\documentclass[12pt]{iopart}
\usepackage{graphicx}
\usepackage{overpic}
\usepackage{amssymb}

\begin{document}
\title[]{Suppression of parametric instabilities in inhomogeneous plasma with multi-frequency light}

\author{Yao Zhao$^{1,\dag}$, Suming Weng$^{2,3}$, Zhengming Sheng$^{2,3,4}$, Jianqiang Zhu$^{1,3}$}

\address{$^1$Key Laboratory of High Power Laser and Physics, Shanghai Institute of Optics and Fine Mechanics, Chinese Academy of Sciences, Shanghai 201800, China}
\address{$^2$Key Laboratory for Laser Plasmas (MoE), School of Physics and Astronomy, Shanghai Jiao Tong University, Shanghai 200240, China}
\address{$^3$Collaborative Innovation Center of IFSA (CICIFSA), Shanghai Jiao Tong University, Shanghai 200240, China}
\address{$^4$SUPA, Department of Physics, University of Strathclyde, Glasgow G4 0NG, UK}

\ead{$^{\dag}$yaozhao@siom.ac.cn}

\begin{abstract}
The development of parametric instabilities in a large scale inhomogeneous plasma with an incident laser beam composed of multiple-frequency components is studied theoretically and numerically. Firstly, theoretical analyses of the coupling between two laser beamlets with certain frequency difference $\delta\omega_0$ for parametric instabilities is presented. It suggests that the two beamlets will be decoupled when $\delta\omega_0$ is larger than certain thresholds, which are derived for stimulated Raman scattering (SRS), stimulated Brillouin scattering (SBS), and two plasmon decay (TPD), respectively. In this case, the parametric instabilities for the two beamlets develop independently and can be controlled at a low level provided the laser intensity for individual beamlet is low enough. Secondly, numerical simulations of parametric instabilities with two or more beamlets ($N\sim20$) have been carried out and the above theory model is validated. Simulations confirm that the development of parametric instabilities with multiple beamlets can be controlled at a low level, provided the threshold conditions for $\delta\omega_0$ is satisfied, even though the total laser intensity is as high as $\sim10^{15}$W/cm$^2$. With such a laser beam structure of multiple frequency components ($N\gtrsim20$) and total bandwidth of a few percentages ($\gtrsim4\%$), the parametric instabilities can be well-controlled.
\end{abstract}


\maketitle

\section{Introduction}

Laser plasma instabilities \cite{Kaw2017,Klimo2010Particle}, especially stimulated Raman scattering (SRS), stimulated Brillouin scattering (SBS) and two-plasmon decay (TPD) instability, are among the critical issues, which could prevent the realization of the inertial confinement fusion (ICF) ignition \cite{Lindl2014Review,craxton2015direct,Betti2016Inertial,Batani2014,Weber2015Temperature}. Therefore, the investigation of the fundamental physics and possible suppression strategies about the laser plasma instabilities is necessary \cite{Sheng2018,ping2019enhanced,baker2018high,rosenberg2018origins}. Many ideas have been proposed to suppress parametric instabilities over the last three decades, such as various beam smoothing techniques \cite{lehmberg1983use,moody2001backscatter}, broadband lasers \cite{thomson1974effects,eimerl2016stardriver,ZHAO2017Stimulated}, and external magnetic field \cite{winjum2018mitigation} etc. More recently, a new type of laser beams called decoupled broadband lasers is proposed\cite{YaoZ2017Effective}. It is made of many frequency components. Under certain conditions, these different components are decoupled and the parametric instabilities can be effectively suppressed. So far, the suppression effect with such laser beam structure is only investigated in homogeneous plasma. In this work, we consider the parametric instabilities control with such a multi-frequency laser beam in a large scale inhomogeneous plasma.

In homogeneous plasma, the laser beamlets with different frequencies can be coupled via Langmuir waves or ion acoustic waves when their instability regions overlap \cite{YaoZ2017Effective}. However, this coupling mechanism is not suitable for inhomogeneous plasmas due to the mismatch of wavenumbers outside the local resonant region. According to the linear model, the instability modes grow in a local region, and gradually saturate after propagating out of the resonant region \cite{rosenbluth1972,liu1974raman}. In this work, we investigate the propagation of multi-frequency light in a large scale inhomogeneous plasma, and give the conditions for the effective suppression of parametric instabilities with a multi-frequency beam. As long as the suppression criterions are satisfied, the hot electron productions and the saturation amplitude of the backscattering light are significantly reduced. The theoretical model is supported by particle-in-cell (PIC) simulations.

\section{Theoretical analysis of the propagation of two light beams in inhomogeneous plasmas}
\subsection{Linear model for convective instability}

Here we consider the spatial amplification of the instability modes in a plasma with density profile $n_e=n_0(1+x/L)$, where $L$ is the density scale length and $x$ is the longitudinal axis. The driving laser beam is composed of many beamlets with different frequencies,
\begin{equation}
a=\sum^N_{i=1}a_i\cos(\omega_it+\phi_i),
\end{equation}
where $a_i$ is the normalized amplitude of $i$-th beamlet with a carrier frequency $\omega_i$ and a random phase $\phi_i$, and $N$ is the number of beamlets. The relation between $a_i$ and laser intensity $I_i$ is given by $a_i=\sqrt{I_i(\mathrm{W}/\mathrm{cm}^2)[\lambda(\mu \mathrm{m})]^2/1.37\times 10^{18}}$. To simplify the problem, we first study the convective instability developed by two light beamlets with different frequencies, i.e., $N=2$. Assuming the two lights have an equal amplitude $a_1=a_2=a_0/\sqrt{2}$, and different frequencies $\omega_1=\omega_0-\delta\omega_0/2$ and $\omega_2=\omega_0+\delta\omega_0/2$, where $\omega_0$ is the central frequency, and $\delta\omega_0$ is the light bandwidth. We have an approximation for $\delta k_0=k_2-k_1\approx\omega_0\delta\omega_0/k_0c^2$ with $k_0$ the wavenumber of central frequency. According to previous studies on the effect of laser bandwidth on the parametric instability in homogeneous plasma \cite{thomson1974effects}, when $\delta\omega_0\gg\Gamma_0$, a modified temporal growth rate is given by $\Gamma_m=\Gamma_0^2/\delta\omega_0$ for the whole incident light. In the following, we study the convective process by using the modified $\Gamma_m$,
\begin{equation}
\nu_sa_s+v_s\partial_xa_s=\frac{\Gamma_m}{\sqrt{2}}a_p\left[\exp\left(i\int K_1dx\right)+\exp\left(i\int K_2dx\right)\right],
\end{equation}
\begin{equation}
\nu_pa_p-v_p\partial_xa_p=\frac{\Gamma_m}{\sqrt{2}}a_s\left[\exp\left(-i\int K_1dx\right)+\exp\left(-i\int K_2dx\right)\right],
\end{equation}
where $a_s$ and $a_p$ respectively are the normalized amplitude of the scattered light and plasma wave, $\nu_s$ and $\nu_p$ are the damping for the scattered light and the plasma wave, respectively. Wavenumber mismatch $K_1=k_1-k_s-k_p=k_0-\delta k_0/2-k_s-k_p$ and $K_2=k_2-k_s-k_p=k_0+\delta k_0/2-k_s-k_p$, where $k_s$ and $k_p$ are the wavenumber for the scattered light and the plasma wave, respectively. The above Eqs. (2) and (3) can be reduced to
\begin{equation}
\nu_sa_s+v_s\partial_xa_s=\sqrt{2}\Gamma_ma_p\exp(iK_0'x^2/2)\cos(\delta k_0'x^2/4),
\end{equation}
\begin{equation}
\nu_pa_p-v_p\partial_xa_p=\sqrt{2}\Gamma_ma_s\exp(-iK_0'x^2/2)\cos(\delta k_0'x^2/4),
\end{equation}
where $K_0=k_0-k_s-k_p$, $K_0'=dK_0/dx$, and $\delta k_0'=d\delta k_0/dx$. Considering a heavy damping for the plasma wave $a_p$, the saturation coefficient is obtained to the first order,
\begin{equation}
G=\frac{2\pi\Gamma_m^2}{v_sv_pK_0'}\left[1+\cos\left(\frac{2\nu_p^2\delta k_0'}{K_0'^2v_p^2}\right)\right].
\end{equation}
Equation (6) indicates that the two different frequency beams can be coupled to develop convective instability in a same resonant region. However, the saturation level is lowered by the bandwidth with comparing to the Rosenbluth gain saturation coefficient. For SRS instability, Eq. (6) can be reduced to
\begin{equation}
G_{SRS}\approx\frac{2\pi\Gamma_m^2}{v_sv_pK_0'}\left[1+\cos\left(\frac{4L\omega_0\nu_p^2\omega_L^2}{ n_0c(\omega_0^2-\omega_{pe}^2)^{3/2}}\delta\omega_0\right)\right],
\end{equation}
where $\omega_L=\sqrt{\omega_{pe}^2+3k_L^2v_{th}^2}$ is the frequency of Langmuir wave with $v_{th}$ being the electron thermal velocity. An approximation can be made based on Eq. (7) that the two beamlets are mutually independent when
\begin{equation}
\delta\omega_0\gtrsim\frac{\pi n_0c(\omega_0^2-\omega_{pe}^2)^{3/2}}{8L\omega_0\omega_L^2\nu_p^2}.
\end{equation}
Considering a plasma with $n_0=0.08n_c$, $T_e=3$keV and $L=3000\lambda$ with $\lambda$ being the central light wavelength in vacuum, and the corresponding Landau damping is $\nu_p\approx0.055\omega_0$ \cite{kruer1988physics}, we have the threshold $\delta\omega_0\approx2.4\%\omega_0$. Consequently, for suppressing SRS, bandwidth between any two beamlets is in the order of $10^{-2}\omega_0$. For the case of $N$ beamlets, the frequency difference between two neighboring beamlets is $\delta\omega_0=\Delta\omega_0/(N-1)$, where $\Delta\omega_0$ is the total bandwidth of the multi-frequency light. Therefore, a small beamlet number $N$ is more suitable for the suppression of SRS when the total bandwidth is in the order of $\Delta\omega_0\sim10^{-2}\omega_0$. Different from the instability-region-coupling in homogeneous plasma, the threshold Eq. (8) is independent of laser amplitude due to the wavenumber detuning out of the local resonant region, which we will discuss in detail in the next section.

\subsection{Secondary amplification of scattering lights}

\begin{figure}
\centering
    \begin{tabular}{lc}
        \begin{overpic}[width=0.6\textwidth]{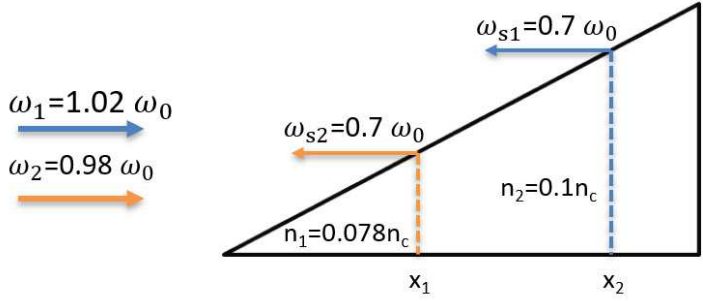}
        \end{overpic}
    \end{tabular}
\caption{ Schematic diagram showing an example for the secondary amplification of backscattering light developed by two laser beamlets at frequencies $\omega_1$ and $\omega_2$. The backscattering light from $x_2$ will be shared at $x_1$.
    }
\end{figure}

Situations are rather complicated for the scattering light propagating in a large scale inhomogeneous plasma. A scattering light produced by one incident light can be amplified again as a seed mode in the subsequent parametric excitation process in a region where its frequency is equal to the scattering light developed by another light. Therefore, the above linear model is suitable for describing the behavior of Langmuir waves, due to its linear propagation in the whole inhomogeneous plasma. Briefly, the bandwidth weakens the strength of longitudinal electrostatic field and therefore the production of hot electrons, however, a part of the scattering light produced by one incident light may be magnified by the other light when the frequencies of the two scattering lights have a cross range. For example, as shown in Fig. 1, the scattering light developed by $\omega_1=1.02\omega_0$ at $0.1n_c$ has frequency $\omega_{s1}=0.7\omega_0$, and it will be amplified again by the other incident light $\omega_2=0.98\omega_0$ as a seed mode at $0.078n_c$, where the frequency of the scattering light is also $\omega_{s2}=0.7\omega_0$. The amplification coefficient can be estimated by
\begin{equation}
G_{SL}=2\pi\left(\frac{\Gamma_1^2}{v_{s1}v_{p1}K_{01}'}+\frac{\Gamma_2^2}{v_{s2}v_{p2}K_{02}'}\right),
\end{equation}
where the subscripts $1$ and $2$ refer to the parameters at $x_1$ and $x_2$, respectively. Therefore, the saturation level of this part of scattered lights has not been greatly reduced due to the secondary amplification. For a finite inhomogeneous plasma with Langmuir wave frequency range $[\omega_{L1},\omega_{L2}]$, the two scattering lights developed by SRS are amplified independently in the propagation when $\omega_2-\omega_{L2}\gtrsim\omega_1-\omega_{L1}$, i.e.,
\begin{equation}
\delta\omega_0>\omega_{L2}-\omega_{L1}.
\end{equation}

Considering an inhomogeneous plasma with density range $[0.08,0.1]n_c$, the threshold for suppression of the secondary amplification of scattering light is $\delta\omega_0>3.3\%\omega_0$. By comparing with the linear threshold Eq. (8), Eq. (10) is relatively larger for a large scale inhomogeneous plasma. Therefore, both the hot electron production and the reflectivity are well-controlled when Eq. (10) is satisfied.

Different from SRS, the frequency of SBS backscattering light changes little with the plasma density. For an inhomogeneous plasma with density $[n_1,n_2]$, the frequency range of the scattering light is $|\delta\omega_s|\approx2c_s\omega_{pe}(\sqrt{n_2}-\sqrt{n_1})/k_0c$. Therefore, the suppression threshold for SBS is
\begin{equation}
\delta\omega_0>2c_s\omega_{pe}(\sqrt{n_2}-\sqrt{n_1})/k_0c\sim10^{-4}\omega_0.
\end{equation}
Note that a small frequency difference $\delta\omega_0\sim10^{-3}\omega_0$ is sufficient for the effective suppression on SBS. The beamlet number is in the order of $N\sim10$ for a multi-frequency light with total bandwidth $\Delta\omega_0=(N-1)\delta\omega_0\sim10^{-2}\omega_0$.

\subsection{Threshold for suppressing two plasmon decay instability}

According to the TPD dispersion relation in cold homogeneous plasmas \cite{kruer1988physics}
\begin{equation}
(\omega^2-\omega_{pe}^2)[(\omega-\omega_0)^2-\omega_{pe}^2]=\Gamma_\mathrm{TPD}^2,
\end{equation}
where $\Gamma_\mathrm{TPD}$ is the temporal growth rate of TPD, we know that TPD is a local instability, and always happens in a narrow region $[0.5\omega_0-\Gamma_\mathrm{TPD},0.5\omega_0+\Gamma_\mathrm{TPD}]$ with $\Gamma_\mathrm{TPD}\approx k_0ca_0/4$. Therefore, the frequency deviation between different beamlets can separate their developing region, and each beam will be independent when
\begin{equation}
\delta\omega_0/\omega_0>\frac{k_0ca_i}{2\omega_0}\sim0.433a_i.
\end{equation}
Note that Eq. (13) has not included the temperature effects. For a beamlet with $a_i\sim10^{-3}$, the threshold for suppressing TPD is around $\delta\omega_0\sim10^{-3}\omega_0$. Therefore, the beamlet number is in the order of $N\sim10$ for a multi-frequency light with total bandwidth $\Delta\omega_0\sim10^{-2}\omega_0$.

In summary, we have presented the required frequency difference between two laser beams for their decoupling. Once they are decoupled, one can simply control the parametric instabilities by controlling a single beamlet. With this, one can design the beam structure for the driving light under a given intensity and bandwidth. In the following section, we will carry out numerical simulation to test the theory predictions found in this section.

\section{Simulations for the suppression of the parametric instabilities in inhomogeneous plasmas}

To validate the above theoretical prediction, a series of particle-in-cell (PIC) simulations have been performed with different bandwidth by using the {\sc klap} code \cite{chen2008development}. Sections 3.1 and 3.2 are devoted to SRS and SBS, where the results are mainly obtained from one-dimensional (1D) simulations. Sec. 3.3 is devoted to TPD, where the results are obtained from two-dimensional (2D) simulations.

\subsection{Suppression of stimulated Raman scattering}

The space and time given in the following are normalized by the laser wavelength $\lambda$ and the laser period $\tau$ in vacuum. The length of the simulation box is 600$\lambda$, where the plasma occupies a region from 50$\lambda$ to 550$\lambda$ with density profile $n_e(x)=0.08[1+(x-50)/1000]n_c$, i.e., the density range for this finite inhomogeneous plasma is $[0.08,0.12]n_c$. The initial electron temperature is $T_{e0}=2$keV. Here we only consider the SRS effects, therefore the ions are immobile with a charge $Z=1$. A linearly-polarized semi-infinite pump lasers with a uniform amplitude $a_0=0.014$ (the corresponding intensity is $I_0=2.5\times10^{15}\mathrm{W}/\mathrm{cm}^2$ with $\lambda=0.33\mu m$) is incident from the left boundary of the simulation box. We have taken 100 cells per wavelength and 50 particles per cell. In the simulation, we change the number of laser beamlets $N$ while keeping the total incident beam energy conserved, and therefore the beamlet amplitude is $a_i=a_0/\sqrt{N}$, i.e., $a_i=0.01$ for $N=2$ and $a_i=0.0031$ for $N=20$.

\begin{figure}
\centering
    \begin{tabular}{lc}
        \begin{overpic}[width=0.98\textwidth]{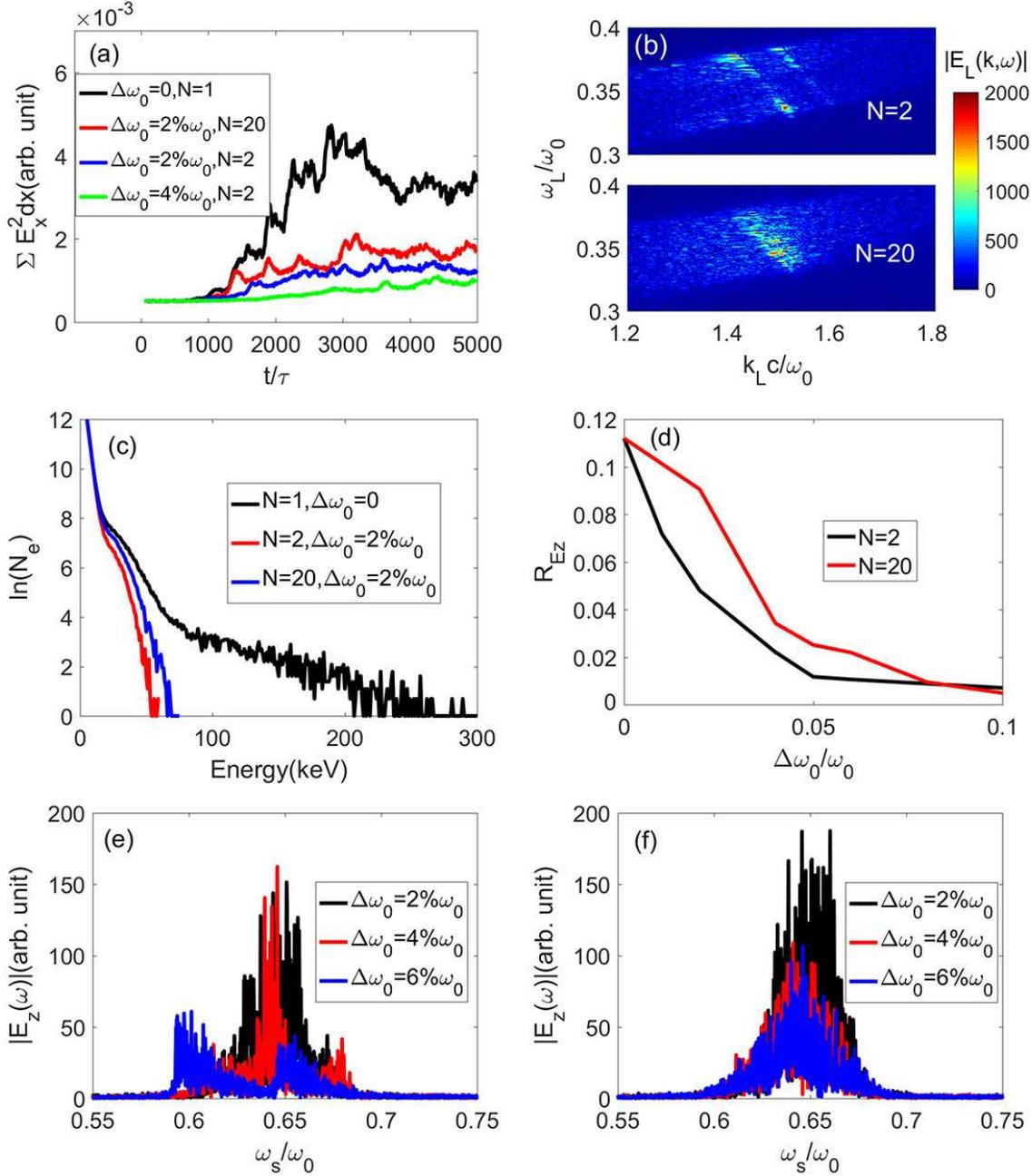}
        \end{overpic}
    \end{tabular}
\caption{ (a) Temporal evolution of electrostatic energy for the incident light with different bandwidths. (b) Distributions of the Langmuir wave in $(k_L,\omega_L)$ space obtained for the time window [1500,2500]$\tau$ with different beam number $N$ under a same bandwidth $\Delta\omega_0=4\%\omega_0$. (c) Energy distributions of electrons found for the normal laser beam and the multi-frequency light with different beam number $N$ under the same energy and bandwidth $\Delta\omega_0=2\%\omega_0$. $N_e$ is the relative electron number. (d) Reflectivity of backscattering light for incident light with different bandwidth and beam number. (e) and (f) Spectra of the backscattering light found, respectively, for the multi-frequency light composed of $N=2$ and $N=20$ beamlets under the same energy.
    }
\end{figure}

As we can see from Fig. 2(a), comparing to the case with a single frequency beam, the beam with finite bandwidth $\Delta\omega_0=2\%\omega_0$ can significantly reduce the strength of the Langmuir wave in the inhomogeneous plasma. The saturation level is lowered further when the bandwidth $\Delta\omega_0$ increases to $4\%\omega_0$. Therefore, with the beam structure proposed in the last section, the bandwidth of incident light can reduce the saturation level of SRS. Comparing two cases in Fig. 2(a), one finds that the suppression effect is weakened when increasing the beamlet number $N$ under a same bandwidth. In homogeneous plasma, the instability region of each beamlet is shrunk by reducing the amplitude $a_i=a_0/\sqrt{N}$ when the total energy is unchanged. Therefore, the beamlets are gradually decoupled with the increase of $N$. Different from this, the suppression condition Eq. (8) in inhomogeneous plasma is independent of the laser amplitude, due to the wavenumber mismatch outside the resonant region. Therefore, the beamlets are still coupled when the beamlet number increased, which will be proved by the following phase plot.

The linear relation between the frequency difference of two beamlets $\delta\omega_0$ and wavenumber difference of Langmuir wave $\delta k_L$ is
\begin{equation}
\delta k_L=(dk_L/d\omega_0)\delta\omega_0=\left(\frac{\omega_0}{\sqrt{\omega_0^2-\omega_{pe}^2}}+\frac{\omega_0-\omega_{pe}}{\sqrt{\omega_0^2-2\omega_{pe}\omega_0}}\right)\delta\omega_0/c.
\end{equation}
For a case with $N=2$ and $\delta\omega_0=\Delta\omega_0=4\%\omega_0$, we have $\delta k_Lc\sim0.087\omega_0$ at $n_e=0.1n_c$. The phase plot presented in Fig. 2(b) indicates that the instability region has already been separated by the frequency difference $\delta\omega_0=4\%\omega_0$ when $N=2$. Under the same conditions, the phases are strongly coupled for relatively small amplitude $a_i=0.0031$ with $\delta\omega_0=\Delta\omega_0/(N-1)\approx0.2\%\omega_0$. Therefore, the phase coupling of incident beamlets has no relations to their amplitudes in inhomogeneous plasmas.

As shown in Fig. 2(c), the electron temperature is reduced by the light with $\Delta\omega_0=2\%\omega_0$, due to the suppression of Langmuir wave discussed above. One finds a long hot-tail heated by absolute SRS via SRS rescattering at $\Delta\omega_0=0$ \cite{zhao2019absolute}. The reduction of first-order SRS leads to the unsatisfied threshold for developing SRS rescattering. The electron temperature for $N=2$ case is slightly lower than the one $N=20$, which is in agreement with the Langmuir wave strength in Fig. 2(a).

Figure 2(d) shows that the reflectivity of backscattering light decreases with the increase of bandwidth. Note that the reflectivity is still large at $\Delta\omega_0=2\%\omega_0$, even though the Langmuir wave has been greatly reduced under the same condition as shown in Fig. 2(a). This is mainly because of the secondary amplification of the scattering lights as discussed in Sec. 2.2. In this example, the range of Langmuir wave frequency is $[0.33,0.38]\omega_0$, the threshold for suppressing secondary amplification is $\delta\omega_0\sim5\%\omega_0$ according to Eq. (10). Considering the case with $\Delta\omega_0=2\%\omega_0$ and $N=2$, the spectra of the two scattering light respectively are $\omega_{s1}=[0.61,0.66]\omega_0$ and $\omega_{s2}=[0.63,0.68]\omega_0$, where an overlapping frequency range $[0.63,0.66]\omega_0$ can be found. Therefore, the scattering light $\omega_{s1}=[0.63,0.68]\omega_0$ will be amplified again as a seed mode when it propagates into the resonant region of $\omega_{s2}=[0.63,0.68]\omega_0$. From Fig. 2(e) we know that the sharing parts are shrunk with the increase of the bandwidth, and beams are totally separated until $\Delta\omega_0=6\%\omega_0$. The scattering lights are strongly coupled for $N=20$ even at $\Delta\omega_0=6\%\omega_0$ as presented in Fig. 2(f). This further proved that the threshold for decoupling of incident beamlets are independent of light amplitude in inhomogeneous plasma.

Generally speaking, the decoupled laser beam structure with certain bandwidth not only reduces the reflectivity and the SRS rescattering process, bus also suppresses the Langmuir wave amplitude. As a result, the hot electron production is also significantly reduced.

\subsection{Suppression of stimulated Brillouin scattering}

\begin{figure}
\centering
    \begin{tabular}{lc}
        \begin{overpic}[width=0.98\textwidth]{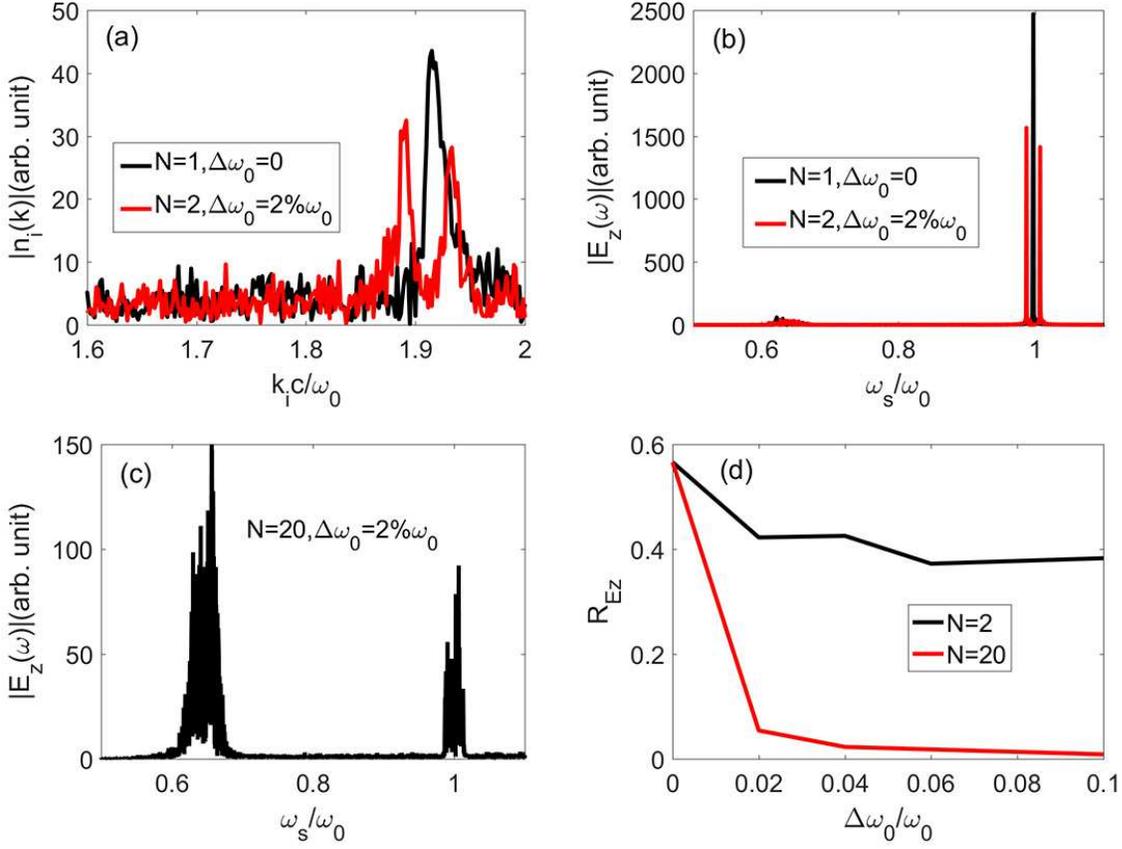}
        \end{overpic}
    \end{tabular}
\caption{ (a) The wave-number distributions for the normal laser beam and the two beamlets with $\Delta\omega_0=2\%\omega_0$ under the same light energy. (b) The corresponding spectra of the backscattering light to (a). (c) Spectra of the backscattering light for the multi-frequency light with $\Delta\omega_0=2\%\omega_0$, and $N=20$. (d) Reflectivity of backscattering light for incident light with different bandwidth and beam number.
    }
\end{figure}

To further validate the suppression effects of bandwidth on SBS, we performed a series of 1D PIC simulations with mobile ions. The mass of ion is $m_i=1836m_e$ with a charge $Z=1$. To develop an intense SBS, we set the ion temperature $T_{i0}=0$ in our simulations. The other parameters are the same as the above simulations.

Different from SRS, a very small bandwidth is sufficient to effectively suppress SBS according to Eq. (11). Therefore, the suppression effect is better for the light with larger beam number $N=20$ under a fixed bandwidth. The wavenumber distributions of ion-acoustic-wave is presented in Fig. 3(a) for two cases with $N=1$ and $N=2$. The instability regions are separated when $\Delta\omega_0=2\%\omega_0$. Even though the amplitude of each beamlet for $N=2$ is smaller than the $N=1$ case, the intensity is still large enough to excite intense SBS. Therefore, a large amount of lights are scattered out of the plasma as shown in Fig. 3(b). Note that SRS (the corresponding spectrum is around $\omega_s\sim0.62\omega_0$) has been greatly suppressed by SBS under this condition \cite{Zhao2017Inhibition}. Figure 3(c) indicates that SBS is significantly suppressed by increasing the beamlet number $N$ to 20 under the same bandwidth $\Delta\omega_0=2\%\omega_0$ with comparing to Fig. 3(b). The frequency difference between two neighbouring beamlets is $\delta\omega_0\approx10^{-3}\omega_0>\delta\omega_s\sim10^{-4}\omega_0$, therefore beamlets are decoupled and each beamlet is too weak to develop intense SBS. SRS is the dominant instability when SBS is suppressed, which indicates that the optimum inhibition parameters are different for SRS and SBS. From Fig. 3(d), we can see that the reflectivity is finally maintained at 30\% for $N=2$, and is well below 4\% for $N=20$.

In conclusion, SRS can be greatly suppressed by a multi-frequency light with a broad bandwidth and small beamlet number $N\lesssim10$. On the contrary, SBS can be effectively inhibited by a small bandwidth and large beamlet number $N\gtrsim20$. As a tradeoff, an effective suppression of both SRS and SBS can be found for the light with bandwidth $\Delta\omega_0\gtrsim4\%\omega_0$ and $N\approx20$.

\subsection{Suppression of two plasmon decay instability}

\begin{figure}
\centering
    \begin{tabular}{lc}
        \begin{overpic}[width=0.98\textwidth]{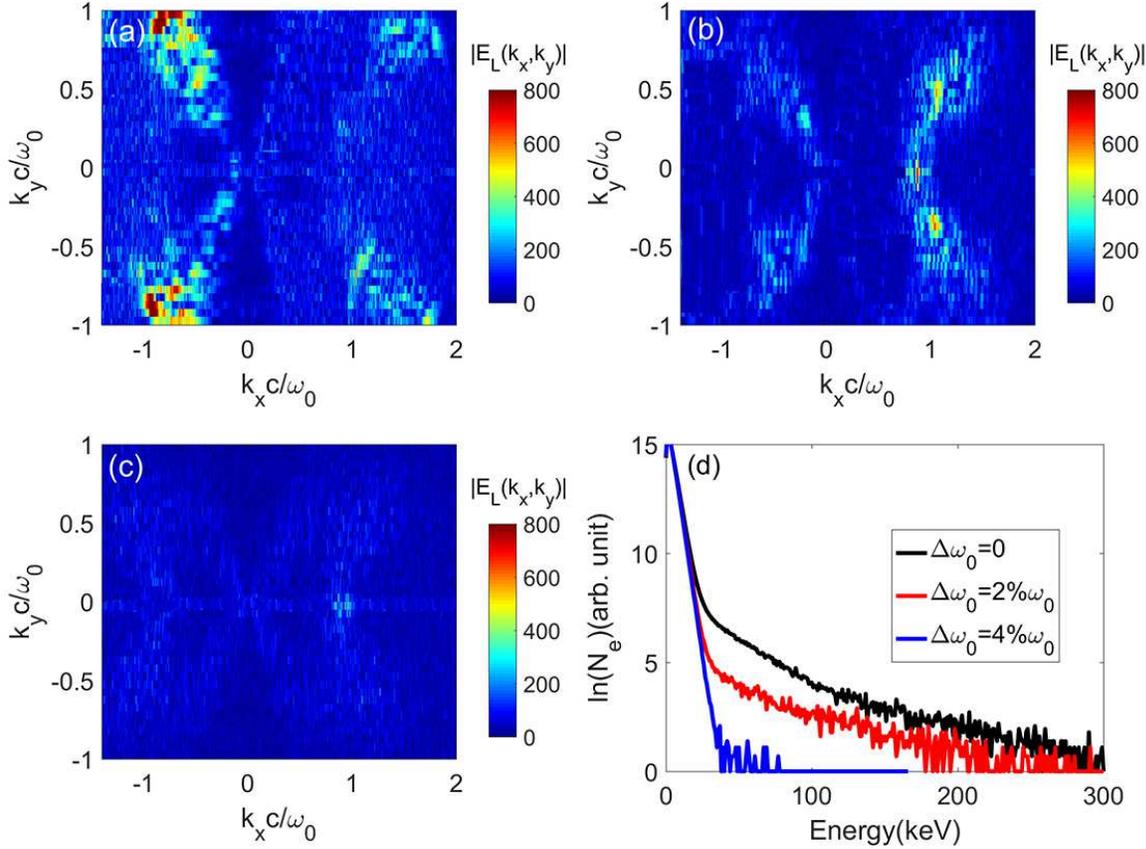}
        \end{overpic}
    \end{tabular}
\caption{ (a), (b) and (c) Spatial Fourier transform $|E_L(k_x,k_y)|$ of the electric field at $t=800\tau$ under $\Delta\omega_0=0$, $\Delta\omega_0=2\%\omega_0$ and $\Delta\omega_0=4\%\omega_0$, respectively. (d) Energy distributions of electrons for different bandwidth at $t=1200\tau$.
    }
\end{figure}

Different from the above two stimulated scattering instabilities, TPD always happens in a local region near $0.25n_c$. To validate the suppression effects on TPD, we have performed several two-dimensional (2D) simulations. The length of the simulation box is 600$\lambda$, where the plasma occupies a region from 30$\lambda$ to 180$\lambda$ with the density profile $n_e(x)=0.22[1+(x-30)/660]n_c$. Here we mainly consider the TPD instability, therefore the ions are immobile with a charge $Z=1$. The initial electron temperature is $T_{e0}=2$keV. A p-polarized (electric field of light is parallel to the simulation plane) semi-infinite pump lasers with a uniform amplitude $a_0=0.014$ is incident from the left boundary of the simulation box.

Similar to SBS, the frequency-difference Eq. (13) for suppressing TPD is in the order of $\sim10^{-3}$. Therefore, larger beamlet number is better for instability inhibition under a same bandwidth. Without the loss of generality, here we take $N=20$ for the broad bandwidth light, i.e., $a_i=0.0031$ and the threshold of bandwidth is $\Delta\omega_0\gtrsim2.6\%\omega_0$ according to Eq. (13). As can be seen from Figs. 4(a) and 4(b), the strength of TPD is reduced by the bandwidth $\Delta\omega_0=2\%\omega_0$. However, TPD is still intense enough to heat abundant electrons as presented in Fig. 4(d). When the threshold is completely satisfied $\Delta\omega_0=4\%\omega_0>2.6\%\omega_0$, TPD is almost totally suppressed, and only a weak SRS mode can be found in Fig. 4(c). From Fig. 4(d), we know that hot electrons are greatly suppressed under $\Delta\omega_0=4\%\omega_0$. The results are consistent with some previous fluid simulations about TPD suppression \cite{follett2018suppressing}.

\section{Summary}

In summary, we have studied theoretically and numerically the suppression of parametric instabilities in a large scale inhomogeneous plasma with a unique laser beam, which is composed of many beamlets, each at a different frequency. The suppression effect occurs when the frequency difference between any two beamlets is larger than certain value, so that there is no coupling between any two beamlets. Approximate thresholds for the required frequency difference are obtained for the effective suppression of SRS, SBS and TPD instabilities. Different from those for the homogeneous plasma case, the thresholds for SRS and SBS in inhomogeneous plasma are independent of the laser amplitude, due to the mismatch of wavenumbers outside the local resonant region. Provided the total bandwidth of the multi-frequency light is in the order of $\Delta\omega_0\sim10^{-2}\omega_0$, a small beamlet number $N\lesssim10$ is more suitable for the suppression of SRS in a large scale inhomogeneous plasma. Comparing with SRS, SBS can be greatly suppressed for laser beams with only a small bandwidth due to a slight change in the scattered-light frequency. Therefore, the optimal parametric ranges are different for the suppression of SRS and SBS. A tradeoff can be made for the bandwidth of the laser beam and the number of beamlets, so that both SRS and SBS can be effectively suppressed. Since TPD always develops near 0.25$n_c$, the beamlets are decoupled when their corresponding instability regions have no overlap, as long as the frequency difference between any two beamlets exceeds certain threshold. Our theoretical model is validated by particle-in-cell simulations, and effective suppressions of reflectivity and hot-electron productions are found at the threshold conditions.

\section{Acknowledgement}

The authors acknowledge useful discussions with C. S. Liu. This work was supported by the Natural Science Foundation of Shanghai (No. 19YF1453200) and the National Natural Science Foundation of China (Nos. 11775144 and 1172109).

\end{document}